\documentclass[prb,twocolumn,showpacs]{revtex4}
\usepackage{graphicx}
\usepackage{amssymb}
\usepackage{amsmath}

\begin{document}

\title{Mesoscopic transition in the shot noise of diffusive S/N/S junctions}

\author{C. Hoffmann}
\altaffiliation[Present address: ]{Institute of Physics, University of Basel, Klingelbergstrasse 82, 4056 Basel, Switzerland}
\email{christian.hoffmann@unibas.ch}
\author{F. Lefloch}
\author{M. Sanquer}
\affiliation{D\'epartement de Recherche Fondamentale sur la Mati\`ere Condens\'ee/SPSMS/LCP\\ CEA Grenoble, 17 rue des Martyrs, 38054 Grenoble cedex 09, FRANCE}

\author{B. Pannetier}
\affiliation{Centre de Recherches sur les Tr\`es Basses Temp\'eratures-C.N.R.S. associated to Universit\'e Joseph Fourier, 25 rue des Martyrs, 38042 Grenoble, FRANCE}

\date{\today}

\begin{abstract}
We experimentally investigated the current noise in diffusive Superconductor/Normal metal/Superconductor junctions with lengths between the superconducting coherence length $\xi_{\Delta}$ and the phase coherence length $L_{\Phi}$ of the normal metal ($\xi_\Delta <L<L_{\Phi}$). We measured the shot noise over a large range of energy covering both the regimes of coherent and incoherent multiple Andreev reflections. The transition between these two regimes occurs at the Thouless energy where a pronounced minimum in the current noise density is observed. Above the Thouless energy, in the regime of incoherent multiple Andreev reflections, the noise is strongly enhanced compared to a normal junction and grows linearly with the bias voltage. Semi-classical theory describes the experimental results accurately, when taking into account the voltage dependence of the resistance which reflects the proximity effect. Below the Thouless energy, the shot noise diverges with decreasing voltage which may indicate the coherent transfer of multiple charges.
\end{abstract}


\maketitle

\begin{figure}[t]
\includegraphics[width=0.48\textwidth,bb= 140 340 700 760,clip]{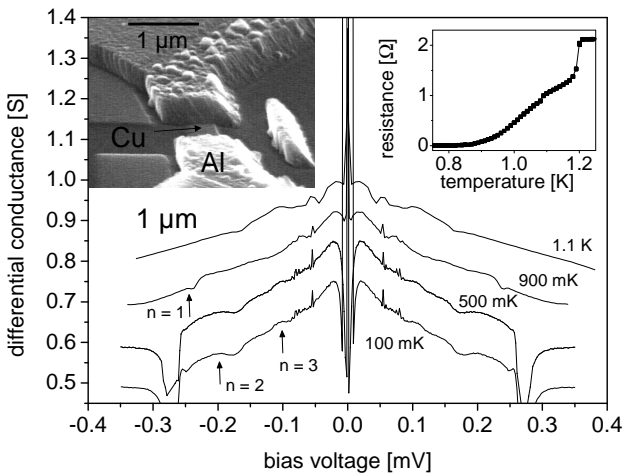}
\caption{Differential conductance $dI/dV$ versus bias voltage at various temperatures for sample l (data are shifted by $0.1, 0.2$ and $0.35\, S$ for $T=500\, mK, 900\, mK$ and $1.1\, K$ respectively). Left inset : scanning electron micrograph of a typical sample. Right inset : resistance of the sample as a function of temperature. The drop at $T = 1.2 \, K$ is due to the superconducting transition of the aluminum reservoirs.}
\label{fig:gv}
\end{figure}

Although the influence of the proximity effect between a superconductor (S) and a normal metal (N) on the conductance of hybrid SN structures has been under study for decades, the impact of the presence of charge pairs on the current noise has been investigated experimentally only recently~\cite{Jehl00,Kozhevnikov00,Lefloch03,Reulet03}. The transport at a SN interface is mediated by Andreev Reflection (AR). An electron with energy $|\epsilon|<\Delta$ with respect to the Fermi level cannot escape from the normal metal into the superconductor due to the absence of electronic states in the gap $\Delta$. Instead, it enters the superconductor together with a second electron to create a Cooper pair and a hole is retroreflected in the normal metal. The electron and the retroreflected hole states are coherent, in the 
diffusive limit, over a distance $L_c=min(L_{\phi},\xi_{\epsilon}=\sqrt{\hbar D/ \epsilon})$ where $D$ is the diffusion constant of the normal metal and $L_{\phi}$ the single particle phase coherence length.\\
In S/N/S junctions with a normal metal length $L>L_c$, the Andreev pair is split up and the electron and the hole behave, far from the interface, as independent quasiparticles. In this {\it incoherent regime} the quasiparticles produce shot noise which originates from the diffusion through the normal metal. The noise is enhanced compared to a N/N/N system because each quasiparticle entering the normal region is successively retroreflected at the two SN interfaces (incoherent multiple Andreev reflections - IMAR). This implies many passages of quasiparticles through the junction, instead of only one in the normal case. These IMAR persist as long as the quasiparticle energy is within the interval $-\Delta < \epsilon < \Delta$ and no inelastic collisions occur. The effect of the inelastic interactions on shot noise in S/N/S junctions has been studied by various groups~\cite{Hoss00,Roche01,Hoffmann02}. It was shown that electron-electron interaction reduces the energy window of accessible states for the quasiparticles participating to the IMAR leading to a decrease of the current noise density.\\
In short S/N/S junctions, the situation is somewhat more complicated. Indeed, as long as $L<L_c$ (that means the Thouless energy $E_{Th}=\hbar D/ L^2$ exceeds the bias voltage \footnote{It should be emphasized at this point that voltage and temperature do not affect the proximity effect the same way. Indeed, the temperature corresponds to a spread in energy of the distribution functions for the electrons and the holes, whereas the voltage corresponds to a shift. Therefore, $L_c$ is not a straight cut-off when dealing with the temperature but should rather be seen as a decay length.}), successive Andreev reflections at the two interfaces are coherent and the interference between quasiparticles leads to the formation of Andreev bound states. In this {\it coherent regime}, the bound states can carry a supercurrent and one observes dc and ac Josephson effects. In very short junctions with a length $L$ smaller than the superconducting coherence length $\xi_{\Delta}=\sqrt{\hbar D/ \Delta}$ (equivalent to $E_{Th}>\Delta$) only two bound states exist and the transport via these states can be considered as the transfer of effective charges $2\Delta /eV$. Then, the noise at low voltage can be interpreted as the shot noise of these effective charges due to Landau-Zener transitions between the bound states~\cite{Naveh99}.\\
A very interesting situation can be reached in S/N/S junctions where $\xi_{\Delta}<L<L_{\Phi}$. In this case, one can tune the transition from the regime of coherent pair transport ($eV<E_{Th}<\Delta$) to single quasiparticle transport ($E_{Th}<eV<\Delta$) by varying the external voltage. In this letter, we present noise measurements in diffusive S/N/S junctions with such an intermediate length. A clear change in the transport mechanisms at the Thouless energy is revealed and appears as a pronounced minimum in the current noise density. This transition, characteristic of the transport in hybrid SN structures at the mesoscopic scale, can be achieved experimentally but remains difficult to explain theoretically. Existing theories concern either diffusive junctions with negligible proximity effect, which are accessible with semi-classical models~\cite{Nagaev01,Bezuglyi01} or the fully coherent situation~\cite{Naveh99}. The latter applies to the noise properties of coherent superconducting atomic point contacts with a small number of conducting channels~\cite{Cuevas99}. In this situation, the experimental results are well understood~\cite{Cron01}. In contrast, the interpretation of the few experimental investigations of noise in multichannel S/N/S junctions available up to now~\cite{Hoss00,Roche01,Strunk02,Hoffmann02} is still a puzzle.\\
In order to measure the current noise, we used a SQUID-based experimental setup~\cite{Jehl99}. The input coil of the SQUID is connected in series with a reference resistor of $0.123\, \Omega$ and the sample. The current fluctuations propagating in this loop are transformed into voltage fluctuations by the SQUID. The intrinsic noise level is about $8\, \mu \Phi_0 /\sqrt{Hz}$ which is equivalent to $1.6\, pA /\sqrt{Hz}$ in the input coil of the SQUID. The noise is measured in the frequency range $10\, Hz$ - $12\, kHz$. At frequencies above $2\, kHz$, 1/f-noise contributions are negligible for all bias currents.\\
The S/N/S junctions are fabricated by shadow evaporation of Cu and Al at different angles through a PMMA-PMMA/MAA bilayer mask in an ultra-high-vacuum chamber. First, a $50\, nm$ thick copper island is evaporated and immediately after, two $480\, nm$ thick aluminum reservoirs. The left inset of figure~\ref{fig:gv} shows a scanning electron micrograph of a typical sample. We studied samples with lengths between $0.4$ and $0.85\, \mu m$ and widths from $0.2$ to $0.4\, \mu m$. The results presented here concern mainly one sample (referred as sample l) with length $0.85\, \mu m$, width $0.4\, \mu m$ and an overlap between the reservoirs and the copper bridge of about $0.3\times 0.4\, \mu m^2$ on each side. The other samples show similar results and will be mentioned for comparison if necessary.\\
To avoid dealing with proximity effect corrections that reduce the resistance of the copper bridge when the reservoirs become superconducting, the resistance $R_N$ of the sample is evaluated from the value above $T_c$ minus the estimated reservoir resistance. For sample l we obtain $R_N=1.75 \,\pm 0.2\, \Omega$. Then, we can estimate the interface resistance by comparison with a second sample half as long as sample l (but with the same width and the same overlap at the reservoirs) and fabricated on the same wafer ($R_N\simeq 1.05\, \Omega)$. This gives an estimation of a total interface resistance $R_B\simeq 0.4\, \Omega$ and a sheet resistance of $0.65\, \Omega$ for copper (diffusion constant $D=80\, cm^2s^{-1}$). The normal resistance of sample l is therefore dominated by the resistance of the copper film ($\simeq 1.35\, \Omega$).\\
As a function of temperature, the zero bias resistance shows a broad transition between $T_c(Al)=1.2\,K$ and $0.8\, K$, below which a supercurrent arises (see right inset of figure \ref{fig:gv}). This behavior is evidence that the phase coherence length $L_{\phi}$ is longer than the sample length, at least for temperatures below $800 \, mK$.
At finite bias, we observe subharmonic gap structures (SGS) marked by a local maximum in the conductance when $eV=2\Delta/n$, see Fig.~\ref{fig:gv}. We can identify peaks for $n=2$ and $3$ over the whole temperature range, whereas the $n=1$ peak is masked for $T<900\, mK$ by a transition, probably induced by the bias current. The origin of the additional peaks at $V\simeq 0.06\, meV$ is not clear.\\
\begin{figure}[h]
\includegraphics[width=0.48\textwidth,bb=0 0 750 545,clip]{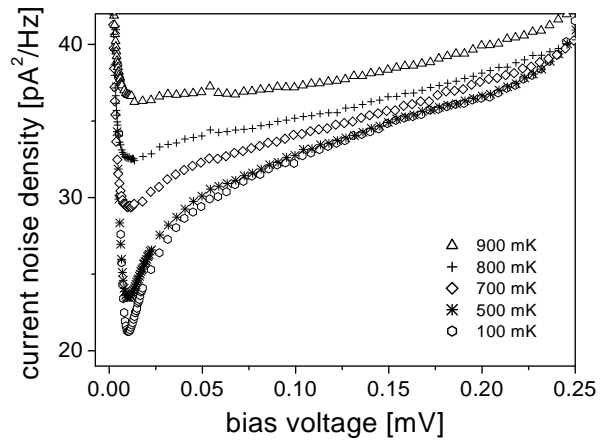}
\caption{Current noise density $S_I$ versus bias voltage at various temperatures. We observe a minimum at $eV\approx E_{Th}$.}
\label{fig:sv}
\end{figure}
The overall shape of the current noise density as a function of bias voltage is shown in Fig.~\ref{fig:sv}. We observe a pronounced minimum at $V= 10\, \mu V$ corresponding roughly to the Thouless energy $E_{Th}\simeq 7\, \mu eV$ of sample l. This minimum indicates the transition from the regime of coherent pair transport to the regime where the Andreev pairs are split up into independent quasiparticles before reaching the opposite interface.\\
The noise behavior at high voltage can be understood within a simplified model. Consider an electron entering the normal metal at the energy $\epsilon \approx -\Delta$. At the first SN-interface it is Andreev-reflected into a hole which travels through the normal region a second time. At the other SN-interface the hole is again retroreflected as an electron and so forth. In the incoherent case the phase information between two subsequent Andreev reflections is lost and no interference is possible. The quasiparticle energy is increased by $eV$ when it travels from one interface to the other. Therefore, the quasiparticle can escape to the superconducting electrodes only after $N$ passages, with $N=2\Delta /(eV)+1$, reaching an energy $\epsilon \gtrsim \Delta$.
\begin{figure}[t]
\includegraphics[width=0.48\textwidth,bb=85 280 810 780,clip]{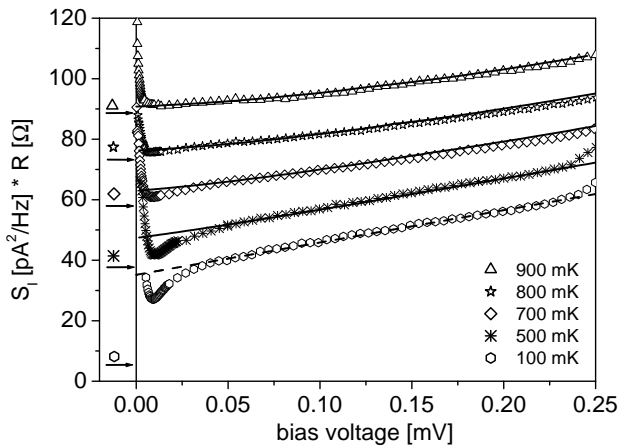}
\caption{Current noise density times the resistance $R=V/I$ versus bias voltage at various temperatures (the data curves are shifted successively by $10\, pA^2/Hz$). The data at $100\, mK$ can be compared to the theoretical prediction in the zero temperature limit (Eq. (\ref{SiV})) with $\Delta=165\mu eV$ (dashed line). At higher temperatures, the thermal noise of quasiparticles outside the gap has to be taken into account. The predictions following Ref.~\onlinecite{Bezuglyi01} are shown as solid lines. The arrows indicate the thermal noise level $4k_BT$ corresponding to each data curve (including the shift).}
\label{fig:srv}
\end{figure}
Within this description, each quasiparticle entering the normal part causes a series of incoherent Andreev reflections which leads to the diffusion of $N$ quasiparticles through the normal part. The total current noise is therefore the shot noise of a diffusive metal $\frac{1}{3}2eI$ times $N$:
\begin{equation}
S_I(V)=\frac{1}{3}2eI\times N=\frac{2}{3R}(eV+2\Delta). \label{SiV}
\end{equation}
This is exactly the prediction of semiclassical theory in the zero temperature limit and in the absence of inelastic processes~\cite{Nagaev01,Bezuglyi01}. In this model, the proximity corrections are neglected. In our junctions however, such corrections persist over the whole voltage range (see Fig.~\ref{fig:gv}). An expression for the noise taking into account the proximity effect has been derived recently in SIN junctions with a tunnel barrier at the interface~\cite{Pistolesi03} but is still lacking in the S/N/S case. In order to include the observed voltage dependence of the resistance, we use $R(V)=V/I$ in equation (\ref{SiV}) rather than the normal state resistance $R_N$ and analyze the product $S_I(V) R(V)$ in Fig.~\ref{fig:srv}.\\
At $T=100\, mK$ we obtain very good agreement between experiment and Eq. (\ref{SiV}) with $\Delta =165\, \mu eV$, shown in Fig.~\ref{fig:srv} by the dashed line, in the range from $50\, \mu V$ up to the current induced transition at about $250\, \mu V$. Up to now, this linear regime of IMAR was only approximately achieved with a large scatter in the data~\cite{Strunk02}. Note that a fit of $S_I(V)$ using a constant resistance instead of the measured $R(V)$, requires unreasonable values $R_N=2.5\, \Omega$ and $\Delta=330\mu eV$.\\
At temperatures above $300\, mK$, the thermal noise of the quasiparticles outside the gap have to be taken into account. Along the lines of Ref.~\onlinecite{Bezuglyi01} we can write the total noise as a sum of this thermal noise and the subgap noise (Eq. (12) and (13) in Ref.~\onlinecite{Bezuglyi01}). The fits obtained using the BCS temperature dependence of the superconducting gap, show excellent agreement with the experimental data between $T=500\, mK$ and $900\, mK$ (solid lines).\\
So far, we considered only the linear part of the noise at high voltage. However, for decreasing voltage ($V<50\, \mu V$) the experimental data show a nonlinear regime which extends down to the minimum at the Thouless energy. The simple model used above to derive Eq.(\ref{SiV}) supposes that the quasiparticles reach the gap without inelastic interactions and the corresponding voltage range is therefore called ``collisionless regime''. However at low voltage and finite temperature, the effective length of the junction for the multiple retroreflected particles $L_{eff}=NL\sim L\Delta/V$ exceeds the inelastic length $L_{in}$. In this ``interacting regime'' e-e-collisions interrupt the successive incoherent multiple Andreev reflections before the quasiparticles reach the gap. In the case of strong interaction a Fermi distribution with an effective temperature $T_e$ is restored and the noise equals the corresponding thermal noise~\cite{Hoss00,Roche01,Hoffmann02,Nagaev01,Bezuglyi01}. Details of the analysis in this regime are published elsewhere~\cite{HoffmannPhD}.\\
\begin{figure}[h]
 \centering
 \includegraphics[width=0.35\textwidth,bb=20 40 570 800,clip]{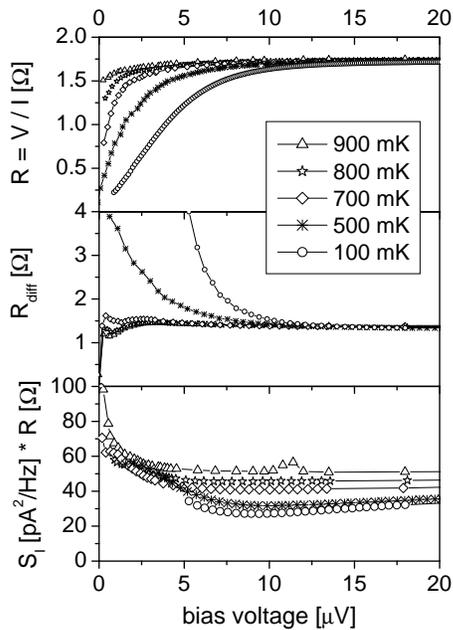}
 \caption{Current noise times resistance, differential resistance and resistance of sample l as a function of the voltage at small bias.}
 \label{fig:datalv}
 \centering
\end{figure}
Decreasing further the voltage below $E_{Th}$, the transport of pairs becomes coherent \cite{Hoss00,Strunk02,Dubos01b}. In this regime, our experimental results reveal a clear increase of the noise as the voltage goes down. We can check that this increase is not due to an equilibrium-like relation $S_I=4k_BT/R(V)$ between the current noise density and the voltage dependent resistance (which indeed decreases near the transition to the dissipationless regime) by considering the product $S_I(V)R(V)$. This is particularly clear in Fig. \ref{fig:datalv} where the low voltage regime is blown up together with the behaviour of the resistance and the differential resistance~\footnote{Note that neither the resistance nor the differential resistance are really constant at $V>10\, \mu V$. Fig. \ref{fig:datalv} gives this impression due to the much smaller voltage range compared to Fig. \ref{fig:gv}.}. It is important to notice that the noise increase persists even at high temperature when the thermal noise level approaches the noise minimum at $eV\approx E_{Th}$ and when the resistance and the differential resistance are almost constant (see curves at $800$ and $900\, mK$ on Fig. \ref{fig:srv} and \ref{fig:datalv}). It is worth noting also that the behavior of the product $S_IR$ at low voltage is almost independent of the temperature whereas the resistance and the differential resistance change significantly. The measurements are complicated by the strong non-linearities at very low temperature and the appearance of hysteresis for $T \leq 300\, mK$. Therefore, at $T=100\, mK$, the noise measurements are restricted to the voltage range $V \geq 5\, \mu V$ where the differential resistance changes not more than a factor 3.
\begin{figure}[h]
 \includegraphics[width=0.22\textwidth,bb=20 280 780 830,clip]{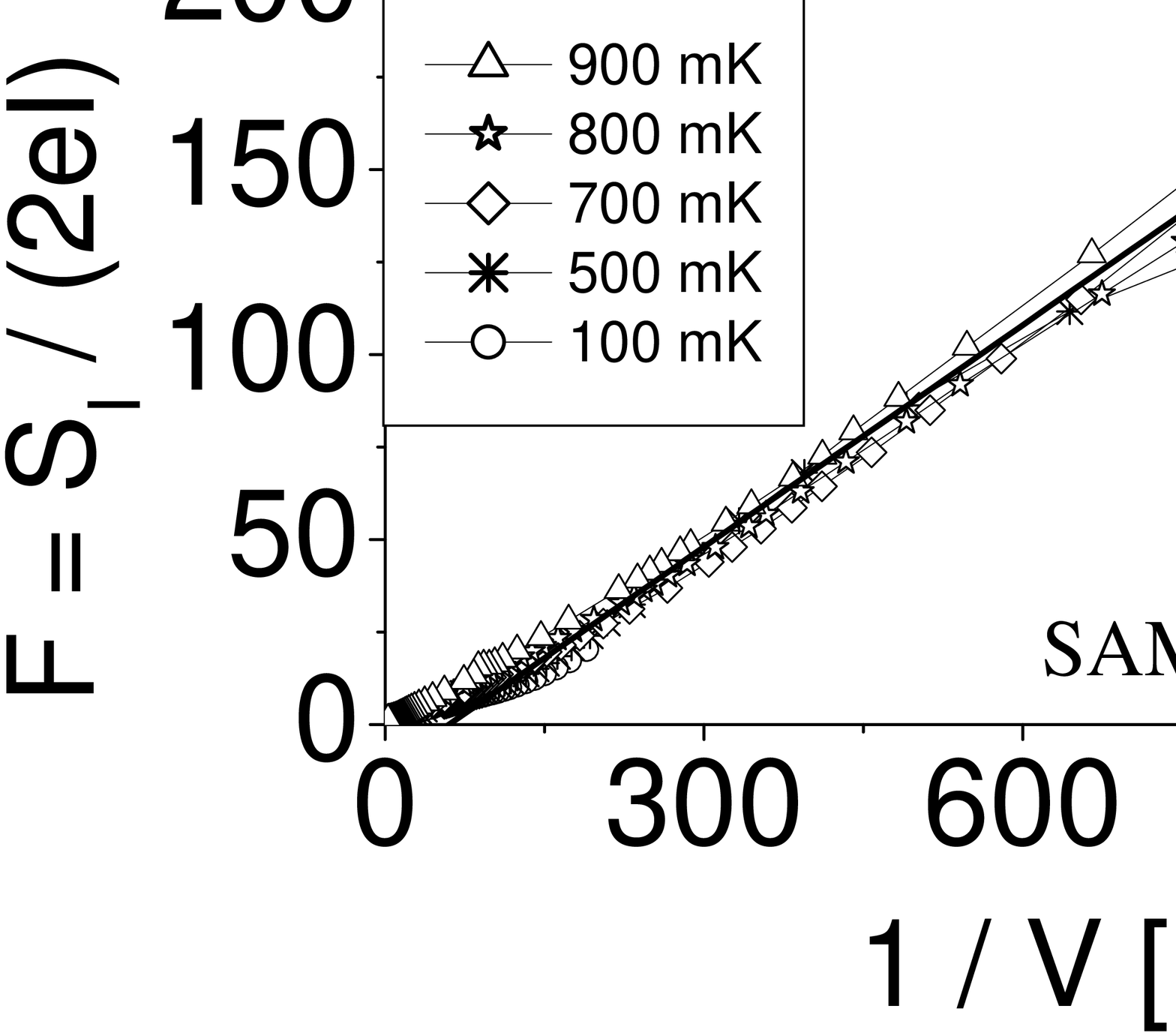}
 \includegraphics[width=0.22\textwidth,bb=0 280 760 830,clip]{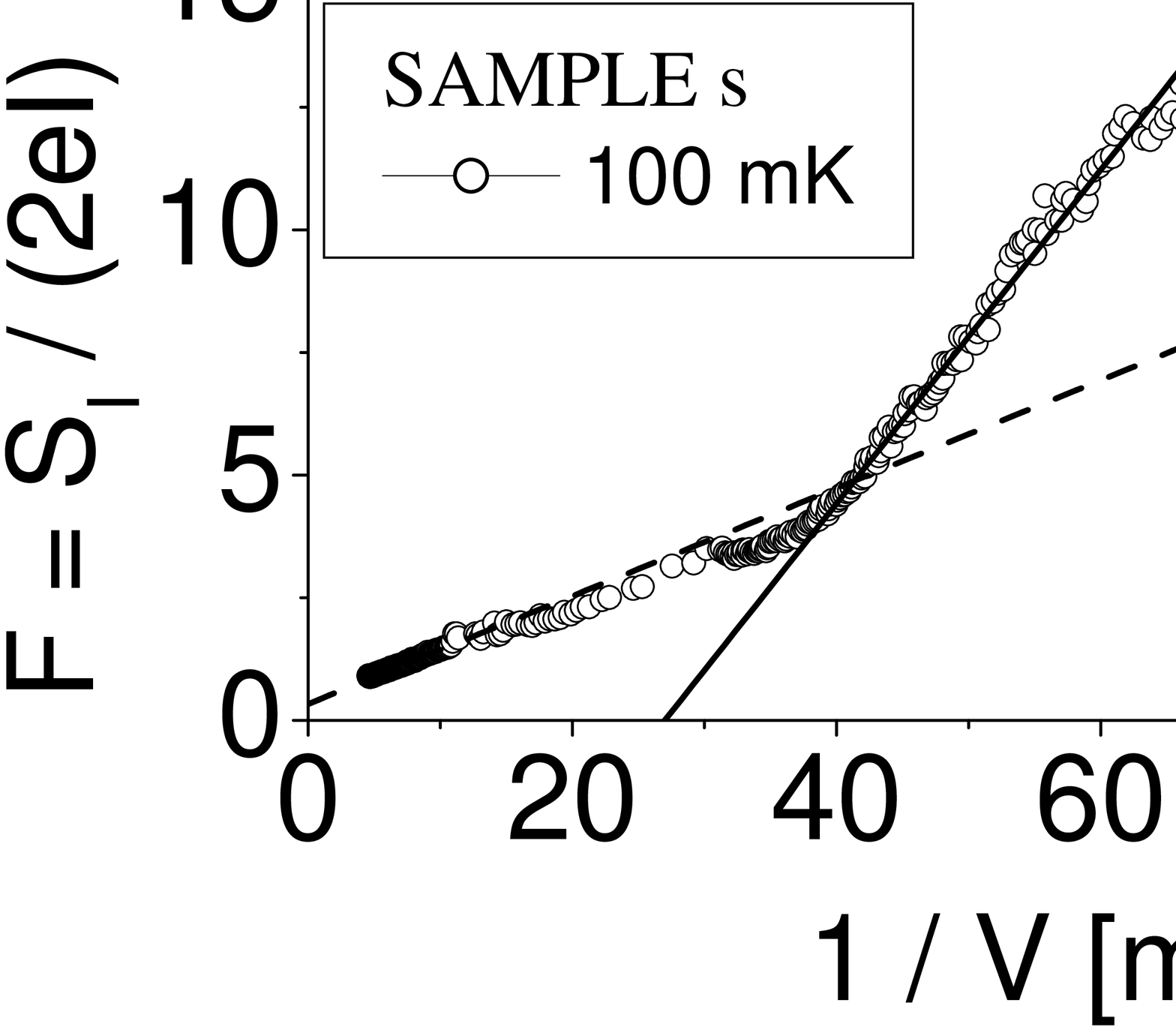}
\caption{Fano factor versus the inverse voltage for two different samples. The Thouless energies are $7\, \mu V$ for sample l (left panel) and $30\, \mu V$ for sample s (right panel). The solid lines have slopes $200\, \mu V$ (left) and $340\, \mu V$ (right). In the right panel, the incoherent regime is also clearly visible (dashed line is a fit according to Eq. (\ref{SiV})).}
 \label{fig:fano}
\end{figure}
Comparison with a shorter sample (width $0.2\, \mu m$ and length $0.4\, \mu m$) is made in Fig. \ref{fig:fano} where the Fano factor $F = S/2eI$ is plotted as a function of the inverse voltage. On the right panel, sample s, the two regimes are distinguishable: above the Thouless energy ($30\, \mu V$ or $1/V < 30\, mV^{-1}$) the Fano factor is linear in $1/V$ with a slope $2\Delta /3e = 110\, \mu V$ as expected for the incoherent regime. Below the Thouless energy the Fano factor is again linear in $1/V$ but with a different slope: $340\, \mu V$. On the left panel, we plot the results obtained on the first sample. The cross-over at the Thouless energy does not clearly appear because the scale is larger than for the shorter sample. However, at voltages below the Thouless energy, the Fano factor is also linear in $1/V$ but with a slope of $200\, \mu V$.
The interpretation of the noise behavior in this low voltage regime is a difficult task because coherent multiple Andreev reflections in the energy window $|\epsilon| \lesssim E_{Th}$ coexist with the diffusion of hot quasiparticles at energies $E_{Th} < |\epsilon| < \Delta$. Moreover, there are no precise calculations for diffusive S/N/S junctions of intermediate size: $\xi_\Delta <L<L_{\Phi}$. In fully coherent quantum point contacts \cite{Cuevas99}, the Fano factor goes like $2\Delta/eV$ and shows a smooth behaviour (no steps) when the transmission in the "N part" is close to 1. In fully coherent ($L << \xi_\Delta$) diffusive SNS junctions \cite{Naveh99}, the Fano factor varies as $F\approx 0.3 (2\Delta/eV)$. Our results reveal that, in diffusive SNS junctions of intermediate size, the Fano factor is also proportional to $1/V$ but with a slope that depends on the length of the sample. However, it is not clear if this behavior is the signature of multiple charge noise as it is the case in superconducting tunnel junctions with pinholes \cite{Dieleman97} and superconducting atomic point contacts \cite{Cron01}. Indications for the presence of coherent MAR in diffusive SNS junctions in the regime $eV<E_{Th}<k_BT$ were found in Ref.~\onlinecite{Dubos01b}. To check if multiple charge noise is present in such junctions, other experiments are required together with theoretical predictions in the appropriate limit ($E_{Th}<\Delta$). In particular, one should include the proximity effect corrections to the density of states of the normal metal. At finite temperature, inelastic interactions will play a role and should appear as a cut-off at low voltage like in the incoherent regime.\\
In conclusion, we present investigations of the shot noise of diffusive S/N/S junctions with intermediate lengths ($\xi_\Delta <L<L_{\Phi}$) which reveal a clear distinction of different transport regimes. At high voltage, the "collisionless regime" is well established and the current noise grows linearly with bias voltage due to incoherent multiple Andreev reflections. The results are in quantitative agreement with semiclassical theory over a large temperature range. A pronounced noise minimum is observed at the Thouless energy. When the applied voltage becomes smaller than the Thouless energy, the Fano factor is found to grow linearly with the inverse voltage but with a slope that depends on the length of the sample. 
\begin{acknowledgments}
We would like to thank M. Houzet, H. Courtois, V. Shumeiko, E. V. Bezuglyi and C. Strunk for fruitful discussions. The samples have been elaborated using the CNRS/NANOFAB clean room facility.
\end{acknowledgments}


\begin{thebibliography}{18}
\expandafter\ifx\csname natexlab\endcsname\relax\def\natexlab#1{#1}\fi
\expandafter\ifx\csname bibnamefont\endcsname\relax
  \def\bibnamefont#1{#1}\fi
\expandafter\ifx\csname bibfnamefont\endcsname\relax
  \def\bibfnamefont#1{#1}\fi
\expandafter\ifx\csname citenamefont\endcsname\relax
  \def\citenamefont#1{#1}\fi
\expandafter\ifx\csname url\endcsname\relax
  \def\url#1{\texttt{#1}}\fi
\expandafter\ifx\csname urlprefix\endcsname\relax\def\urlprefix{URL }\fi
\providecommand{\bibinfo}[2]{#2}
\providecommand{\eprint}[2][]{\url{#2}}

\bibitem[{\citenamefont{Jehl et~al.}(2000)\citenamefont{Jehl, Sanquer,
  Calemczuk, and Mailly}}]{Jehl00}
\bibinfo{author}{\bibfnamefont{X.}~\bibnamefont{Jehl}},
  \bibinfo{author}{\bibfnamefont{M.}~\bibnamefont{Sanquer}},
  \bibinfo{author}{\bibfnamefont{R.}~\bibnamefont{Calemczuk}},
  \bibnamefont{and} \bibinfo{author}{\bibfnamefont{D.}~\bibnamefont{Mailly}},
  \bibinfo{journal}{Nature} \textbf{\bibinfo{volume}{405}}, \bibinfo{pages}{50}
  (\bibinfo{year}{2000}).

\bibitem[{\citenamefont{Kozhevnikov et~al.}(2000)\citenamefont{Kozhevnikov,
  Schoelkopf, and Prober}}]{Kozhevnikov00}
\bibinfo{author}{\bibfnamefont{A.~A.} \bibnamefont{Kozhevnikov}},
  \bibinfo{author}{\bibfnamefont{R.~J.} \bibnamefont{Schoelkopf}},
  \bibnamefont{and} \bibinfo{author}{\bibfnamefont{D.~E.}
  \bibnamefont{Prober}}, \bibinfo{journal}{Phys. Rev. Lett.}
  \textbf{\bibinfo{volume}{84}}, \bibinfo{pages}{3398} (\bibinfo{year}{2000}).

\bibitem[{\citenamefont{Lefloch et~al.}(2003)\citenamefont{Lefloch, Hoffmann,
  Sanquer, and Quirion}}]{Lefloch03}
\bibinfo{author}{\bibfnamefont{F.}~\bibnamefont{Lefloch}},
  \bibinfo{author}{\bibfnamefont{C.}~\bibnamefont{Hoffmann}},
  \bibinfo{author}{\bibfnamefont{M.}~\bibnamefont{Sanquer}}, \bibnamefont{and}
  \bibinfo{author}{\bibfnamefont{D.}~\bibnamefont{Quirion}},
  \bibinfo{journal}{Phys. Rev. Lett.} \textbf{\bibinfo{volume}{90}},
  \bibinfo{pages}{067002} (\bibinfo{year}{2003}).

\bibitem[{\citenamefont{Reulet et~al.}(2003)\citenamefont{Reulet, Kozhevnikov,
  Prober, Belzig, and Nazarov}}]{Reulet03}
\bibinfo{author}{\bibfnamefont{B.}~\bibnamefont{Reulet}},
  \bibinfo{author}{\bibfnamefont{A.~A.} \bibnamefont{Kozhevnikov}},
  \bibinfo{author}{\bibfnamefont{D.~E.} \bibnamefont{Prober}},
  \bibinfo{author}{\bibfnamefont{W.}~\bibnamefont{Belzig}}, \bibnamefont{and}
  \bibinfo{author}{\bibfnamefont{Y.}~\bibnamefont{Nazarov}},
  \bibinfo{journal}{Phys. Rev. Lett.} \textbf{\bibinfo{volume}{90}},
  \bibinfo{pages}{066601} (\bibinfo{year}{2003}).

\bibitem[{\citenamefont{Hoss et~al.}(2000)\citenamefont{Hoss, Strunk,
  Nussbaumer, Huber, Staufer, and Sch{\"o}nenberger}}]{Hoss00}
\bibinfo{author}{\bibfnamefont{T.}~\bibnamefont{Hoss}},
  \bibinfo{author}{\bibfnamefont{C.}~\bibnamefont{Strunk}},
  \bibinfo{author}{\bibfnamefont{T.}~\bibnamefont{Nussbaumer}},
  \bibinfo{author}{\bibfnamefont{R.}~\bibnamefont{Huber}},
  \bibinfo{author}{\bibfnamefont{U.}~\bibnamefont{Staufer}}, \bibnamefont{and}
  \bibinfo{author}{\bibfnamefont{C.}~\bibnamefont{Sch{\"o}nenberger}},
  \bibinfo{journal}{Phys. Rev. B} \textbf{\bibinfo{volume}{62}},
  \bibinfo{pages}{4079} (\bibinfo{year}{2000}).

\bibitem[{\citenamefont{Roche et~al.}(2001)\citenamefont{Roche, Perrin,
  Glattli, Takayanagi, and Akazaki}}]{Roche01}
\bibinfo{author}{\bibfnamefont{P.}~\bibnamefont{Roche}},
  \bibinfo{author}{\bibfnamefont{H.}~\bibnamefont{Perrin}},
  \bibinfo{author}{\bibfnamefont{D.~C.} \bibnamefont{Glattli}},
  \bibinfo{author}{\bibfnamefont{H.}~\bibnamefont{Takayanagi}},
  \bibnamefont{and} \bibinfo{author}{\bibfnamefont{T.}~\bibnamefont{Akazaki}},
  \bibinfo{journal}{Physica C} \textbf{\bibinfo{volume}{352}},
  \bibinfo{pages}{73} (\bibinfo{year}{2001}).

\bibitem[{\citenamefont{Hoffmann et~al.}(2002)\citenamefont{Hoffmann, Lefloch,
  and Sanquer}}]{Hoffmann02}
\bibinfo{author}{\bibfnamefont{C.}~\bibnamefont{Hoffmann}},
  \bibinfo{author}{\bibfnamefont{F.}~\bibnamefont{Lefloch}}, \bibnamefont{and}
  \bibinfo{author}{\bibfnamefont{M.}~\bibnamefont{Sanquer}},
  \bibinfo{journal}{Eur. Phys. J. B} \textbf{\bibinfo{volume}{29}},
  \bibinfo{pages}{629} (\bibinfo{year}{2002}).

\bibitem[{\citenamefont{Naveh and Averin}(1999)}]{Naveh99}
\bibinfo{author}{\bibfnamefont{Y.}~\bibnamefont{Naveh}} \bibnamefont{and}
  \bibinfo{author}{\bibfnamefont{D.~V.} \bibnamefont{Averin}},
  \bibinfo{journal}{Phys. Rev. Lett.} \textbf{\bibinfo{volume}{82}},
  \bibinfo{pages}{4090} (\bibinfo{year}{1999}).

\bibitem[{\citenamefont{Nagaev}(2001)}]{Nagaev01}
\bibinfo{author}{\bibfnamefont{K.~E.} \bibnamefont{Nagaev}},
  \bibinfo{journal}{Phys. Rev. Lett.} \textbf{\bibinfo{volume}{86}},
  \bibinfo{pages}{3112} (\bibinfo{year}{2001}).

\bibitem[{\citenamefont{Bezuglyi et~al.}(2001)\citenamefont{Bezuglyi, Bratus,
  Shumeiko, and Wendin}}]{Bezuglyi01}
\bibinfo{author}{\bibfnamefont{E.~V.} \bibnamefont{Bezuglyi}},
  \bibinfo{author}{\bibfnamefont{E.~N.} \bibnamefont{Bratus}},
  \bibinfo{author}{\bibfnamefont{V.~S.} \bibnamefont{Shumeiko}},
  \bibnamefont{and} \bibinfo{author}{\bibfnamefont{G.}~\bibnamefont{Wendin}},
  \bibinfo{journal}{Phys. Rev. B} \textbf{\bibinfo{volume}{63}},
  \bibinfo{pages}{100501} (\bibinfo{year}{2001}).

\bibitem[{\citenamefont{Cuevas et~al.}(1999)\citenamefont{Cuevas,
  Mart{\'\i}n-Rodero, and Yeyati}}]{Cuevas99}
\bibinfo{author}{\bibfnamefont{J.~C.} \bibnamefont{Cuevas}},
  \bibinfo{author}{\bibfnamefont{A.}~\bibnamefont{Mart{\'\i}n-Rodero}},
  \bibnamefont{and} \bibinfo{author}{\bibfnamefont{A.~L.}
  \bibnamefont{Yeyati}}, \bibinfo{journal}{Phys. Rev. Lett.}
  \textbf{\bibinfo{volume}{82}}, \bibinfo{pages}{4086} (\bibinfo{year}{1999}).

\bibitem[{\citenamefont{Cron et~al.}(2001)\citenamefont{Cron, Goffmann, Esteve,
  and Urbina}}]{Cron01}
\bibinfo{author}{\bibfnamefont{R.}~\bibnamefont{Cron}},
  \bibinfo{author}{\bibfnamefont{M.~F.} \bibnamefont{Goffmann}},
  \bibinfo{author}{\bibfnamefont{D.}~\bibnamefont{Esteve}}, \bibnamefont{and}
  \bibinfo{author}{\bibfnamefont{C.}~\bibnamefont{Urbina}},
  \bibinfo{journal}{Phys. Rev. Lett.} \textbf{\bibinfo{volume}{86}},
  \bibinfo{pages}{4104} (\bibinfo{year}{2001}).

\bibitem[{\citenamefont{Strunk and Sch{\"o}nenberger}(2002)}]{Strunk02}
\bibinfo{author}{\bibfnamefont{C.}~\bibnamefont{Strunk}} \bibnamefont{and}
  \bibinfo{author}{\bibfnamefont{C.}~\bibnamefont{Sch{\"o}nenberger}}, in
  \emph{\bibinfo{booktitle}{Proceedings of the NATO ARW on Quantum Noise in
  Mesoscopic Physics}} (\bibinfo{year}{2002}).

\bibitem[{\citenamefont{Jehl et~al.}(1999)\citenamefont{Jehl, Payet-Burin,
  Baraduc, Calemczuk, and Sanquer}}]{Jehl99}
\bibinfo{author}{\bibfnamefont{X.}~\bibnamefont{Jehl}},
  \bibinfo{author}{\bibfnamefont{P.}~\bibnamefont{Payet-Burin}},
  \bibinfo{author}{\bibfnamefont{C.}~\bibnamefont{Baraduc}},
  \bibinfo{author}{\bibfnamefont{R.}~\bibnamefont{Calemczuk}},
  \bibnamefont{and} \bibinfo{author}{\bibfnamefont{M.}~\bibnamefont{Sanquer}},
  \bibinfo{journal}{Rev. Sci. Instr.} \textbf{\bibinfo{volume}{70}},
  \bibinfo{pages}{2711} (\bibinfo{year}{1999}).

\bibitem[{\citenamefont{Pistolesi et~al.}(2004)\citenamefont{Pistolesi, Bignon, and
  Hekking}}]{Pistolesi03}
\bibinfo{author}{\bibfnamefont{F.}~\bibnamefont{Pistolesi}},
  \bibinfo{author}{\bibfnamefont{G.}~\bibnamefont{Bignon}}, \bibnamefont{and}
  \bibinfo{author}{\bibfnamefont{F.~W.~J.} \bibnamefont{Hekking}},
  \bibinfo{journal}{Phys. Rev. B} \textbf{\bibinfo{volume}{69}},
  \bibinfo{pages}{214518} (\bibinfo{year}{2004}).

\bibitem[{\citenamefont{Hoffmann}(2003)}]{HoffmannPhD}
\bibinfo{author}{\bibfnamefont{C.}~\bibnamefont{Hoffmann}}, Ph.D. thesis,
  \bibinfo{school}{University Joseph Fourier, Grenoble I}
  (\bibinfo{year}{2003}).

\bibitem[{\citenamefont{Dieleman et~al.}(1997)\citenamefont{Dieleman, Bukkems,
  Klapwijk, Schicke, and Gundlach}}]{Dieleman97}
\bibinfo{author}{\bibfnamefont{P.}~\bibnamefont{Dieleman}},
  \bibinfo{author}{\bibfnamefont{H.~G.} \bibnamefont{Bukkems}},
  \bibinfo{author}{\bibfnamefont{T.~M.} \bibnamefont{Klapwijk}},
  \bibinfo{author}{\bibfnamefont{M.}~\bibnamefont{Schicke}}, \bibnamefont{and}
  \bibinfo{author}{\bibfnamefont{K.~H.} \bibnamefont{Gundlach}},
  \bibinfo{journal}{Phys. Rev. Lett.} \textbf{\bibinfo{volume}{79}},
  \bibinfo{pages}{3486} (\bibinfo{year}{1997}).

\bibitem[{\citenamefont{Dubos et~al.}(2001)\citenamefont{Dubos, Courtois,
  Buisson, and Pannetier}}]{Dubos01b}
\bibinfo{author}{\bibfnamefont{P.}~\bibnamefont{Dubos}},
  \bibinfo{author}{\bibfnamefont{H.}~\bibnamefont{Courtois}},
  \bibinfo{author}{\bibfnamefont{O.}~\bibnamefont{Buisson}}, \bibnamefont{and}
  \bibinfo{author}{\bibfnamefont{B.}~\bibnamefont{Pannetier}},
  \bibinfo{journal}{Phys. Rev. Lett.} \textbf{\bibinfo{volume}{87}},
  \bibinfo{pages}{206801} (\bibinfo{year}{2001}).

\end{thebibliography}
\end{document}